\documentclass[preprint,authoryear,12pt]{elsarticle}

\usepackage{graphicx}
\usepackage{color}
\usepackage{amsmath}
\usepackage{amssymb}
\usepackage{bm}
\usepackage{latexsym}
\usepackage{marvosym}
\usepackage{enumerate}
\usepackage{textcomp}
\usepackage{subfigure}
\usepackage{float}
\usepackage{times}
\usepackage[varg]{txfonts}

\usepackage[ps2pdf,%
a4paper=true,%
breaklinks=true,%
colorlinks=true,%
pdfauthor={First Author et al.},%
pdftitle={Kelvin--Helmholtz instability in a twisting solar polar coronal hole jet observed by \emph{SDO}/AIA}%
]{hyperref}

\journal{Advances in Space Research}

\begin{document}

\begin{frontmatter}



\title{Kelvin--Helmholtz instability in a twisting solar polar coronal hole jet observed by \emph{SDO}/AIA}


\author{I.~Zhelyazkov$^1$\corref{cor}}
\address{$^1$Faculty of Physics, Sofia University, 1164 Sofia, Bulgaria}
\cortext[cor]{Corresponding author}
\ead{izh@phys.uni-sofia.bg}


\author{T.~V.~Zaqarashvili$^{2,3,4}$ and L.~Ofman$^{5,6,7}$}
\address{$^2$IGAM, Institute of Physics, University of Graz, 8010 Graz, Austria}
\address{$^3$Abastumani Astrophysical Observatory at Ilia State University, 0162 Tbilisi, Georgia}
\address{$^4$Space Research Institute, Austrian Academy of Sciences, 8042 Graz, Austria}
\address{$^5$Catholic University of America, Washington, DC 20064, USA}
\address{$^6$NASA Goddard Space Flight Center, Greenbelt, MD 20771, USA}
\address{$^7$Visiting, Department of Geosciences, Tel Aviv University, Ramat-Aviv, Tel Aviv 69978, Israel}

\author{R.~Chandra$^{8}$}
\address{$^8$Department of Physics, DSB Campus, Kumaun University, Nainital 263\,001, India}

\begin{abstract}
We investigate the conditions under which the fluting ($m = 2$), $m = 3$, and $m = 12$ magnetohydrodynamic (MHD) modes in a uniformly twisted flux tube moving along its axis become unstable in order to model the Kelvin--Helmholtz (KH) instability in a twisting solar coronal hole jet near the northern pole of the Sun.  We employed the dispersion relations of MHD modes derived from the linearized MHD equations.  We assumed real wavenumbers and complex angular wave frequencies, namely complex wave phse velocities.   The dispersion relations were solved numerically at fixed input parameters (taken from observational data) and varying degrees of torsion of the internal magnetic field.  It is shown that the stability of the modes depends upon five parameters: the density contrast between the flux tube and its environment, the ratio of the external and internal axial magnetic fields, the twist of the magnetic field lines inside the tube, the ratio of transverse and axial jet's velocities, and the value of the Alfv\'en Mach number (the ratio of the tube axial velocity to Alfv\'en speed inside the flux tube).  Using a twisting jet of 2010 August 21 by \emph{SDO}/AIA and other observations of coronal jets we set the parameters of our theoretical model and have obtained that in a twisted magnetic flux tube of radius of $9.8$~Mm, at a density contrast of $0.474$ and fixed Alfv\'en Mach number of ${\cong}0.76$, for the three MHD modes there exist instability windows whose width crucially depends upon the internal magnetic field twist.  It is found that for the considered modes an azimuthal magnetic field of $1.3$--$1.4$~G (computed at the tube boundary) makes the width of the instability windows equal to zero, that is, it suppress the KH instability onset.  On the other hand, the times for developing KH instability of the $m = 12$ MHD mode at instability wavelengths between $15$ and $12$~Mm turn out to be in the range of $1.9$ to $4.7$~minutes that is in agreement with the growth rates estimated from the temporal evolution of the observed unstable jet's blobs in their initial stage.
\end{abstract}

\begin{keyword}
magnetohydrodynamics \sep waves \sep instabilities \sep twisting coronal hole jets \sep numerical methods
\end{keyword}

\end{frontmatter}

\parindent=0.5 cm

\section{Introduction}
\label{sec:intro}
Rotational motion seems to be a common property of various kinds of jets and prominences in the solar atmosphere detected from multi-wavelength observations with high spatial resolution and high cadence using Atmospheric Imaging Assembly (AIA)  \citep{lemen2012}, on board the \emph{Solar Dynamics Observatory\/} (\emph{SDO}) \citep{pesnell2012}, the \emph{Interface Region Imaging Spectrograph\/} (\emph{IRIS}) \citep{depontieu2014}, and the \emph{Hinode\/} \citep{kosugi2007} Extreme-ultraviolet Imaging Spectrometer (EIS) \citep{culhane2007} alongside the earth-basing THEMIS and the Swedish 1-meter solar telescopes.  Solar rotating and helical structures are termed as solar tornadoes.  In fact the word `tornado' was initially associated with solar prominences (see, e.g., \citet{pettit1932} and \citet{schmieder2017} and references therein), but \citet{pike1998} used the same term to describe transition region macrospicules seen by the \emph{Solar and Heliospheric Observatory\/} (\emph{SOHO}) \citep{domingo1995} which (the spicules) have no relation with prominences.  \citet{kamio2010} on using \emph{Hinode}/EIS and the Solar Ultraviolet Measurements of Emitted Radiation instrument (SUMER) \citep{wilhelm1995} on the \emph{SOHO\/} were able to measure the line of sight (LOS) motions of both macrospicule and coronal jets.  At the same time, with the help of the X-Ray Telescope (XRT) \citep{golub2007} on \emph{Hinode\/} and Sun--Earth Connection Coronal and Heliospheric Investigation (SECCHI) instrument suite \citep{howard2008} on the \emph{Solar TErrestrial RElations Observatory\/} (\emph{STEREO}) \citep{kaiser2008} the authors traced the evolution of the coronal jet and the macrospicule.  \citet{wedemeyer2012} performing observations with AIA on board \emph{SDO\/} and the Crisp Imaging Spectropolarimeter (CRISP) \citep{scharmer2003} at the Swedish 1-m Solar Telescope discovered a swirling motion at different heights in the solar atmosphere.  These swirls, also dubbed `magnetic tornadoes' \citep{wedemeyer2013}, the authors found to originate in the chromosphere, but do not appear to be related to any filamentary structure.  These structures can play an important role for channeling energy from the chromosphere into the corona.

In addition to macrospicules, rotational motions have been observed in the so cold Type II spicules \citep{depontieu2012}.  Soft X-ray jets can also exhibit rotational motions.  \citet{moore2013} exploring 54 polar X-ray jets from movies taken by the X-ray Telescope on \emph{Hinode\/} and in the He \mbox{\large \textsc{ii}} $304$~\AA\ band of the \emph{SDO}/AIA have obtained rotational speeds of the order of $60$~km\,s$^{-1}$.  Recently \citet{moore2015} studied $14$ large solar jets that erupted in polar coronal holes and were observed in the outer corona beyond $2.2$\,$R_\odot$ in images from the \emph{SOHO}/Large Angle Spectroscopic Coronagraph (LASCO) \citep{brueckner1995}.  There is no surprise that rotational motions were detected in EUV solar jets, too.  \citet{shen2011} have presented an observational study of the kinematics and fine structure of an unwinding polar jet, with high temporal and spatial observations taken by the \emph{SDO}/AIA and the Solar Magnetic Activity Research Telescope.  In a similar way, \citet{chen2012} using the multi-wavelength data from the \emph{SDO}/AIA, studied a jet occurring in a coronal hole near the northern pole of the Sun and have obtained jet's parameters of the same orders as those in \citet{shen2011} study.  \citet{zhang2014} using the multi-wavelength observations in the EUV passbands from the \emph{SDO}/AIA have detected the onset of jet eruption coinciding with the start time of a C1.6 solar flare.  A rotating coronal hole jet observed with \emph{Hinode\/} and the \emph{SDO}/AIA on 2011 February 8 at around 21:00 UT was reported by \citet{young2014}.  Recently, \citet{filippov2015} analyzed multi-wavelength and multi-viewpoint observations with \emph{STEREO}/SECCHI/EUVI and \emph{SDO}/AIA of a helically twisted plasma jet formed during a confined filament eruption on 2013 April {10--11}.

It is well established that in magnetically structured solar atmosphere various jets with axial mass flow like spicules, surges, EUV and X-ray jets can become unstable against the so called Kelvin--Helmholtz instability (KHI)---for reviews see, e.g., \citet{zhelyazkov2015}, \citet{nakariakov2016}, \citet{zhelyazkov2017} and the references therein.  Recall that the instability exhibits itself as a vortex sheet evolving near jet's boundary which (the vortex sheet) can become unstable to the spiral-like perturbations at small spatial scales provided that jet's axial velocity exceeds some critical/threshold value \citep{ryu2000}.

Previous studies devoted to the KHI modeling in rotating cylindrical jets were carried out by \citet{bodo1989,bodo1996}.  These authors studied the stability of a rotating, magnetized cylindrical axial flow of radius $a$ through an ambient unmagnetized medium by considering that all perturbations of the velocity $\boldsymbol{v}$, magnetic field $\boldsymbol{B}$, and pressure $p$ obey the basic equations of ideal magnetohydrodynamics for a polytropic fluid and are in the form $f(r)\exp[\mathrm{i}(-\omega t + kz + m\theta)]$.  Having derived a Bessel equation for the pressure perturbation $p_1$ and appropriate expression for the radial $\boldsymbol{v}_1$-component the authors merge the solutions in both media via the boundary conditions for continuity of the total (thermal $+$ magnetic) pressure and the Lagrangian displacement (the ratio of radial velocity perturbation component and the angular frequency in the corresponding medium) at the interface $r = a$ and obtain the dispersion relation of the normal MHD modes propagating along the jet.  In their two papers, \citet{bodo1989,bodo1996} have studied analytically and numerically the stability conditions of both axisymmetric, $m = 0$ \citep{bodo1989}, and non-axisymmetric, $|m| \geqslant 1$ \citep{bodo1996}, modes.  A step forward was the study of \citet{zaqarashvili2015}, who examining the stability/instability status or rotating jets, modeling them as moving untwisted/twisted magnetized flux tubes embedded in a homogeneous background magnetic field, have considered the case when the flow velocity is also twisted, thus generalizing in incompressible plasma approximation the wave dispersion equation derived by \citet{bodo1989}.  To finish our survey on the KH modeling in rotating solar jets we should mention the articles of \citet{terradas2008} and \citet{soler2010} who explored the nonlinear instability of kink oscillations in a coronal loop (modeled as an untwisted rotating line-tied magnetic flux tube) due to shear motions and KHI in coronal untwisted magnetic flux tubes due to pure azimuthal shear flows.

In this paper, we study the conditions under which propagating high-mode ($m \geqslant 2$) MHD waves along the polar coronal hole rotating jet might become unstable against the KHI.  In particular, we focus on a test case motivated by recent \emph{SDO}/AIA observation of a twisting jet on 2010 August 21 \citep{chen2012} and on other observations to set the parameters of our analytical model.  In the next section, we provide the details of the observational motivation for the kinematics of the rotating jet.  The geometry of the problem, equilibrium magnetic field configuration and governing equations alongside a concise derivation of the wave dispersion relation are specified in the third section.  Section~\ref{sec:dispers} deals with the numerical solution of the dispersion equation for the $m = 2$, $3$, and $12$ MHD modes and with discussing the obtained results.  In the last section, we summarize the main findings in our research.

\section{Observational motivation for a jet detected near the northern pole of the Sun}
\label{sec:observations}

\begin{figure}[!ht]
   \centering
   \includegraphics[height=.27\textheight]{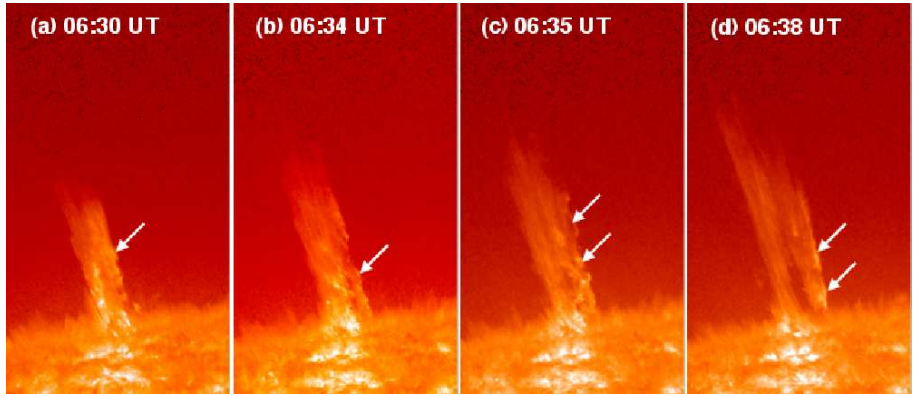}
   \caption{AIA $304$~\AA\ images showing the detailed evolution of the jet.  The small moving blobs on the right side boundary of the jet as indicated by white arrows, could be produced by a KHI.}
   \label{fig:fig1}
\end{figure}
\begin{figure}[!ht]
  \centering
\subfigure{\includegraphics[height=.27\textheight]{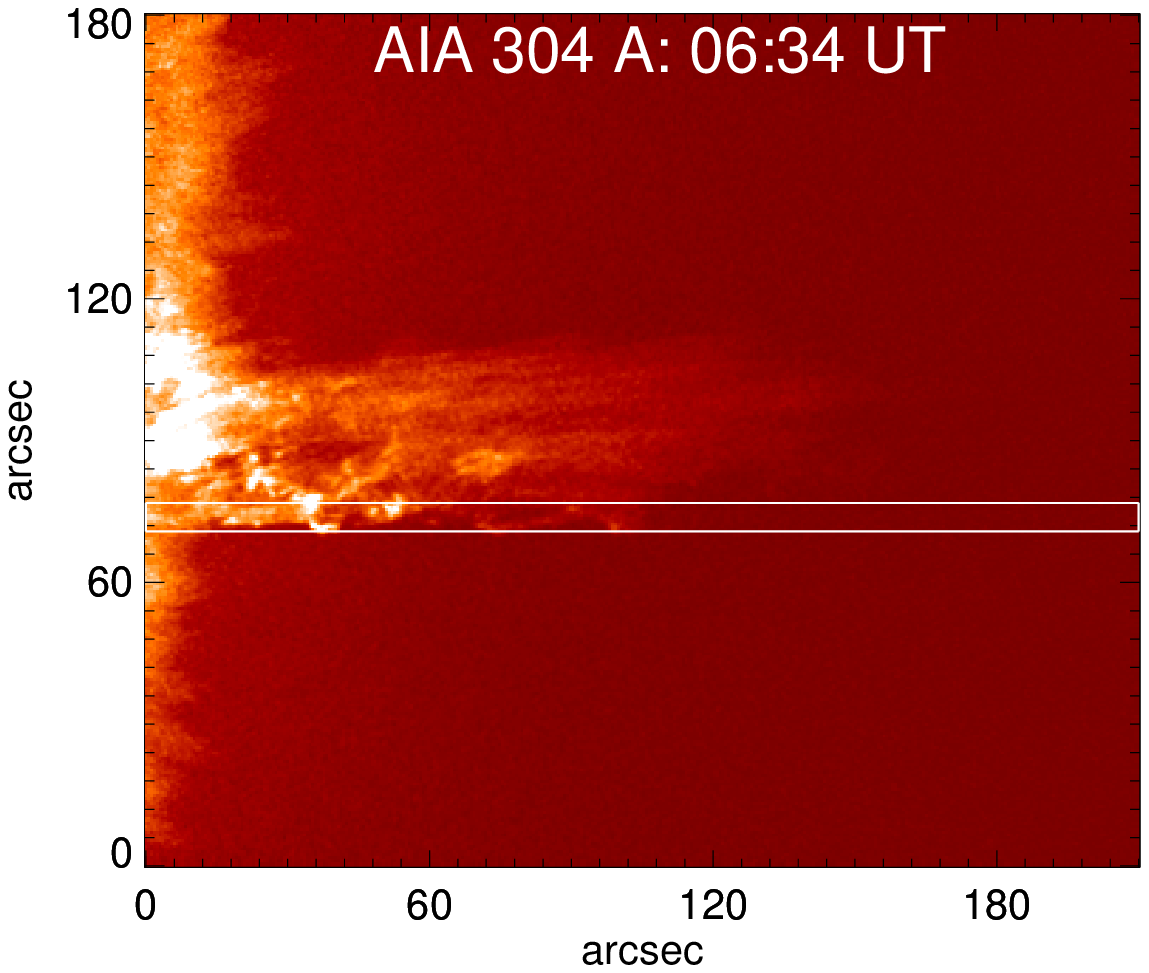}}
\subfigure{\includegraphics[height=.27\textheight]{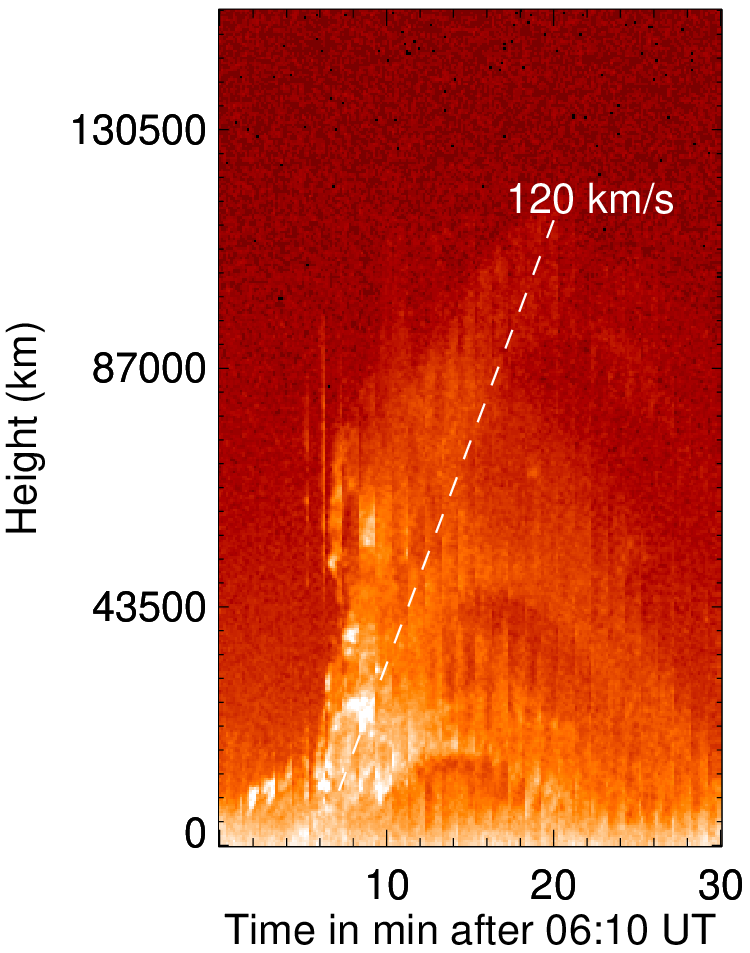}}
  \caption{(Left) AIA $304$~\AA\ image showing the location of slit, where the time slice is taken. The image is rotated to 105 deg for the computation of time slice.  (Right) Time--distance map profile of the jet centered at the blobs location.}
\vspace*{-2mm}
   \label{fig:fig2}
\end{figure}
For the current study, we have selected the jet event of 2010 August 21 as a guide to study the growth of KHI.  This event occurred in the coronal hole region, close to the north pole of the Sun. The jet's limb location provides us the opportunity to calculate the speed with smaller projection effect and reduced line-of-sight integration effect compared to events on the disk.  The event was observed by AIA on board \emph{SDO\/} in different wavelengths.  The pixel size and the temporal resolution of AIA data are $0.60$~arcsec and $12$~s, respectively.  Figure~1 presents the jet's evolution in AIA $304$~\AA.  The jet started around  06:07 UT, reached its maximum height around 06:40 UT.  During the evolution of the jet between 06:32 and 06:38 UT, small scale moving blobs appear on the right boundary.  We interpret these blobs, shown by arrows in Fig.~\ref{fig:fig1}, as evidence of KHI.  In addition to Fig.~\ref{fig:fig1}, we also have attached the AIA $304$~\AA\ movie, where one can see the moving blobs clearly.  To investigate the kinematics of these blobs, we have created the time--distance map along the blobs motion. The location of slit selected for time--distance analysis is given in Fig.~2.  We have computed the speed of the jet/blobs and found ${\sim}120 \pm8$~km\,s$^{-1}$.  Our calculated speed is consistent with the jet speed reported by \citet{chen2012}.  In the time--distance map, we can also notice the downfall of jet material.  The speed of this downfall is ${\sim}60 \pm 7$ km\,s$^{-1}$.

This jet was previously studied by \citet{shen2011} and \citet{chen2012}.
By tracking six identified moving features in the jet, \citet{chen2012} found that the plasma moved at an approximately constant speed along the jet's axis.  Meanwhile, the moving features made a circular motion in the plane transverse to the axis.  Inferred from linear and trigonometric fittings to the axial and transverse heights of six tracks, the authors have found that the mean values of the axial velocity, $U_z$, transfer/rotational velocity, $U_\phi$, angular speed, $\Omega$, rotation period, $T$, and rotation radius, $a$, are $114$~km\,s$^{-1}$, $136$~km\,s$^{-1}$, $0.81^\circ$s$^{-1}$ (or $14.1 \times 10^{-3}$~rad\,s$^{-1}$), $452$~s, and $9.8 \times 10^3$~km, respectively.  \citet{chen2012} on assuming that the magnetic flux across the transverse section of the jet would remain constant, have found that the magnetic field inside the jet gradually decreases with the height from $15 \pm 4$ to about $3 \pm 1$~G at a height of $7 \times 10^4$~km.  These authors do not provide any data concerning the basic physical parameters (electron number density $n$ and electron temperature $T$) of the jet and its environment.  Based on measurements of $n$ and $T$ for similar coronal hole EUV, and X-ray jets, we assume that electron number density is typically $n_\mathrm{jet} = 1.0 \times 10^9$~cm$^{-3}$, and electron temperature is $T_\mathrm{jet} = 1.6$~MK, while the same quantities for the environment are respectively $n_\mathrm{cor} = 0.9 \times 10^9$~cm$^{-3}$ and $T_\mathrm{cor} = 1.0$~MK.  We consider the background magnetic field to be equal to $3$~G---our choice of the magnetic field is based on the evaluation of \citet{pucci2013} who recommend $B = 2.8$~G at $n_\mathrm{jet} = 8 \times 10^8$~cm$^{-3}$ and $T_\mathrm{jet} = 1.6$~MK for standard X-ray jets, while \citet{chandrashekhar2014} claim that $B = 1.21 \pm 0.2$~G perfectly fits the bill for a coronal hole jet with $n_\mathrm{jet} = (5.6$--$6.3)\times 10^8$~cm$^{-3}$ and $T_\mathrm{jet} = 780\,000$--$933\,000$~K.  Thus, we find that the specified above values for the electron number densities, electron temperatures, and magnetic field are consistent with the measured by \citet{chen2012} jet's axial and rotational speeds and we use these parameters in our theoretical model, described below.  Prior to starting the building of our model we have, however, to specify what kind of media are the jet and its environment, more specifically what are their plasma betas.  We accept the \citet{paraschiv2015} suggestion that plasma density in the region crossed by coronal jets, has to be the sum $n_\mathrm{cor} + n_\mathrm{jet}$, that is, in our case equal to $1.9 \times 10^9$~cm$^{-3}$.  Under this assumption, on using a pressure balance equation, we obtain that sound and Alfv\'en speeds in the jet and surrounding plasma are equal to $148$~km\,s$^{-1}$ and ${\cong}150$~km\,s$^{-1}$, and $117$~km\,s$^{-1}$ and $218.0$~km\,s$^{-1}$, respectively.  These speeds yield that jet's plasma beta is equal to ${\cong}1.2$, while that of its environment is just $0.35$.  In their seminal paper on nearly incompressible MHD fluids \citet{zank1993} claim that for regimes with $\beta \sim 1$ and $\beta \ll 1$ the leading-order description of the medium is reduced to two dimensions in the plane perpendicular to the mean magnetic field.  It was shown that the MHD perturbations (considered as nearly incompressible ones) for $\beta \sim 1$ consist of magnetosonic and sound waves with a proper 3D geometry, while Alfv\'en waves should propagate parallel to the magnetic field. For the $\beta \ll 1$ regime, there is a strong tendency for nearly incompressible perturbations to propagate in a 1D direction parallel to the magnetic field.  Thus, in our study for KHI development of Alfv\'en-like perturbations/waves in a one-dimensional modeled rotating solar jet it is appropriate to consider the rotating jet as incompressible plasma and its environment as a cool, also incompressible, medium.

\section{Geometry, the governing equations, and the wave dispersion relation}
\label{sec:basic}
We model the twisted eruption as moving and rotating magnetic flux tube with radius $a$.  The magnetic field inside the tube, $\boldsymbol{B}_\mathrm{i}$, and the rotating velocity, $\boldsymbol{U}$, are twisted, that is,
\begin{equation}
\label{eq:twist}
    \boldsymbol{B}_\mathrm{i} = (0, B_{\mathrm{i}\phi}(r),B_{\mathrm{i}z}) \qquad \mbox{and} \qquad \boldsymbol{U} = (0, U_\phi(r), U_z).
\end{equation}
For simplicity we suppose that the jet has homogeneous density, $\rho_\mathrm{i} = \mathrm{const}$, constant axial component of the magnetic field, $B_{\mathrm{i}z}$, and constant axial velocity, $U_z$.  The rotational jet velocity at the tube boundary, $U_\phi (a) \equiv U_\phi$, determined from observations, in rigid rotation case, can be expressed through the jet angular velocity, $\Omega$, and tube radius, $a$, by the simple relation $U_\phi = \Omega a$.  The jet is assumed to be surrounded by a static homogeneous medium with $\rho_\mathrm{e} = \mathrm{const}$ and uniform axial magnetic field, $(0, 0, B_\mathrm{e})$.

The pressure balance condition inside the jet, derived from integration of the momentum equation for the equilibrium variables, yields the following radial profile of the total pressure:
\begin{equation}
\label{eq:pressurebalance}
    p_\mathrm{t}(r) =  p_\mathrm{t}(0) - \frac{1}{\mu}\int\displaylimits_0^r \frac{B_{\mathrm{i}\phi}^2(s)}{s}\mathrm{d}s + \rho_\mathrm{i} \int\displaylimits_0^r \frac{U_\phi^2(s)}{s} \mathrm{d}s,
\end{equation}
where $\mu$ is the plasma permeability and $p_\mathrm{t}(0)$ is the total (thermal $+$ magnetic) pressure.

Linearized ideal MHD equations, which govern the incompressible dynamics of perturbations in the rotating jet are
\begin{equation}
\label{eq:momentum}
    \frac{\partial}{\partial t}\boldsymbol{v} + (\boldsymbol{U}\cdot \nabla)\boldsymbol{v} + (\boldsymbol{v}\cdot \nabla) \boldsymbol{U} = -\frac{\nabla p_\mathrm{tot}}{\rho_\mathrm{i}} + \frac{\left( \boldsymbol{B}_\mathrm{i} \cdot \nabla \right)\boldsymbol{b}}{\rho_\mathrm{i} \mu} + \frac{\left( \boldsymbol{b} \cdot \nabla \right)\boldsymbol{B}_\mathrm{i}}{\rho_\mathrm{i} \mu},
\end{equation}
\begin{equation}
\label{eq:induct}
	\frac{\partial}{\partial t}\boldsymbol{b} - \nabla \times \left( \boldsymbol{v}
    \times \boldsymbol{B}_\mathrm{i} \right) - \nabla \times \left( \boldsymbol{U} \times \boldsymbol{b}
    \right) = 0,	
\end{equation}
\begin{equation}
\label{eq:divv}
	\nabla \cdot \boldsymbol{v} = 0,	
\end{equation}
\begin{equation}
\label{eq:divb}
	\nabla \cdot \boldsymbol{b} = 0,
\end{equation}
where $\boldsymbol{v} = (v_r, v_\phi, v_z)$ and $\boldsymbol{b} = (b_r, b_\phi, b_z)$ are the perturbations of fluid velocity and magnetic field, respectively, and $p_\mathrm{tot}$ is the perturbation of the total pressure $p_\mathrm{t}$.  We note that the same set of equations with thermal pressure $p = 0$ and $v_z = 0$ will be used for describing the fluid and magnetic field perturbations in the cool jet's environment.  We are not going to re-derive in detail the wave dispersion relation of normal MHD modes propagating in a rotating and axially moving twisted magnetic flux tube---that has been done by \citet{zaqarashvili2015} assuming that surrounding medium is also incompressible magnetized plasma---we will only give the final form of that equation with slightly modified external part for cool medium:
\begin{eqnarray}
\label{eq:dispeq}
    \frac{\left( \sigma^2 - \omega_\mathrm{Ai}^2 \right)F_m(\kappa_\mathrm{i}a) - 2m\left( \sigma \Omega + A\omega_\mathrm{Ai}/\! \sqrt{\mu \rho_\mathrm{i}} \right)}{\rho_\mathrm{i}\left( \sigma^2 - \omega_\mathrm{Ai}^2 \right)^2 - 4\rho_\mathrm{i}\left( \sigma \Omega + A\omega_\mathrm{Ai}/\! \sqrt{\mu \rho_\mathrm{i}} \right)^2} \nonumber \\
    \nonumber \\
    {}= \frac{P_m(\kappa_\mathrm{e} a)}{\rho_\mathrm{e}\left( \sigma^2 - \omega_\mathrm{Ae}^2 \right) - \left( \rho_\mathrm{i}\Omega^2 - A^2/\mu \right)P_m(\kappa_\mathrm{e} a)},
\end{eqnarray}
where
\[
    F_m(\kappa_\mathrm{i}a) = \frac{\kappa_\mathrm{i}aI_m^{\prime}(\kappa_\mathrm{i}a)}{I_m(\kappa_\mathrm{i}a)} \quad \mbox{and} \quad P_m(k_z a) = \frac{\kappa_\mathrm{e} aK_m^{\prime}(\kappa_\mathrm{e} a)}{K_m(\kappa_\mathrm{e} a)}.
\]
Here, the prime means differentiation with respect to the function argument,
\[
    \kappa_\mathrm{i}^2 = k_z^2\left[ 1 - 4\left( \frac{\sigma \Omega + A\omega_\mathrm{Ai}/\!\sqrt{\mu \rho_\mathrm{i}}}{\sigma^2 - \omega_\mathrm{Ai}^2} \right)^2 \right] \quad \mbox{and} \quad \kappa_\mathrm{e}^2 = k_z^2 \left[ 1 - \left( \omega/\omega_\mathrm{Ae} \right)^2 \right]
\]
are the squared wave amplitude attenuation coefficients in both media, in which
\[
    \omega_\mathrm{Ai} = \left( \frac{m}{r}B_{\mathrm{i}\phi} + k_z B_{\mathrm{i}z} \right)/\sqrt{\mu \rho_\mathrm{i}} \quad \mbox{and} \quad \omega_\mathrm{Ae} = k_z B_\mathrm{e}/\sqrt{\mu \rho_\mathrm{e}}
\]
are the corresponding local Alfv\'en frequencies, and
\[
    \sigma = \omega - \frac{m}{r}U_{\phi} - k_z U_z
\]
is the Doppler-shifted wave frequency in the jet.

In numerically solving Eq.~(\ref{eq:dispeq}) we assume that the wave frequency is complex, Re($\omega$) $+$ i\,Im($\omega$), as well as normalize all velocities with respect to the Alfv\'en speed inside the jet, $v_\mathrm{Ai} = B_{\mathrm{i}z}/\sqrt{\mu \rho_\mathrm{i}}$, and all lengths to the tube radius, $a$.  Thus, our dimensionless variables are: normalized wave phase velocity Re($v_\mathrm{ph}/v_\mathrm{Ai}$), wave growth rate Im($v_\mathrm{ph}/v_\mathrm{Ai}$), and wavenumber $k_z a$.  The input parameters are: the density contrast $\eta$, the ratio of external to internal axial magnetic field components $b$, the magnetic field twist parameter $\varepsilon_1$, the flow velocity twist parameter $\varepsilon_2$, and the Alfv\'en Mach number $M_\mathrm{A}$, defined on the next line as
\[
    \eta = \rho_\mathrm{e}/\rho_\mathrm{i}, \quad b = B_\mathrm{e}/B_{\mathrm{i}z}, \quad \varepsilon_1 = Aa/B_{\mathrm{i}z}, \quad \varepsilon_2 = \Omega a/U_z, \quad \mbox{and} \quad M_\mathrm{A} = U_z/v_\mathrm{Ai}.
\]
In the next Sect.~\ref{sec:dispers}, we explore numerically the dispersion characteristics and growth rates of the normal MHD modes traveling on the rotating soft X-ray jet, occurred on the northeast limb (E0N81) of the Sun.

The basic jet's and its environment sound and Alfv\'en speeds, as noted at the end of the Sect.~\ref{sec:intro}, have been determined by the total pressure balance equation (\ref{eq:pressurebalance}) rewritten in the form
\begin{equation}
\label{eq:pbeq}
    p_\mathrm{i} - \frac{1}{2}\rho_\mathrm{i}U_\phi^2 + \frac{B_\mathrm{i}^2}{2\mu} = p_\mathrm{e} + \frac{B_\mathrm{e}^2}{2\mu}.
\end{equation}

For the values of plasma densities $\rho_\mathrm{i}= 1.9 \times 10^9$~cm$^{-3}$ and $\rho_\mathrm{e}= 0.9 \times 10^9$~cm$^{-3}$ (that is, at density contrast $\eta = 0.474$), with $T_\mathrm{i} \equiv T_\mathrm{jet} = 1.6$~MK and $T_\mathrm{e} \equiv T_\mathrm{cor} = 1.0$~MK at jet's rotating speed $U_\phi = \Omega a = 136$~km\,s$^{-1}$ and background magnetic field $B_\mathrm{e} = 3$~G, the above equation yields $B_\mathrm{i} = 2.36$~G, and accordingly, $b = 1.27$.  Measured jet's rotating and axial speeds gives the value of the velocity twist parameter $\varepsilon_2 = 1.2$.  The rest two input parameters, the magnetic field twist parameter $\varepsilon_1$ and Alfv\'en Mach number $M_\mathrm{A}$ will be specified additionally.

\section{Numerical solutions and wave dispersion diagrams}
\label{sec:dispers}
Prior to starting numerical task for solving the dispersion equation (\ref{eq:dispeq}), it is convenient to have some idea at which conditions one can expect the occurrence of KHI.  As shown in \citet{zaqarashvili2015}, at small sub-Alfv\'enic axial jet's speeds, the instability in an untwisted rotating flux tube will occur if
\[
    \frac{a^2 \Omega^2}{v_\mathrm{Ai}^2} > \frac{1 + \eta}{1 + |m|\eta}\,\frac{(k_z a)^2}{|m| - 1}(1 + b^2).
\]
To get some hint for which range of normalized wavenumbers $k_z a$ the instability will occur, we rearrange this inequality to the form
\begin{equation}
\label{eq:instcond}
    (k_z a)_\mathrm{rhs} < \left[ \left( \frac{U_\phi}{v_\mathrm{Ai}} \right)^2 \frac{1 + |m|\eta}{1 + \eta}\,\frac{|m| - 1}{1 + b^2} \right]^{1/2},
\end{equation}
where, recall, $U_\phi = \Omega a$ is the rotating velocity of the jet.

Above inequality shows that the dimensionless wavenumber that limits the instability range in the one-dimensional $k_z a$-space, depends, in addition to jet's parameters $U_\phi$, $v_\mathrm{Ai}$, $\eta$, $b$, also on the wave mode number, $m$.  Moreover, it is clear that only the higher modes, $|m| \geqslant 2$, can have finite instability $k_z a$-ranges.

On the other hand, the jet width, $\Delta \ell$, and its height define a critical dimensionless wavenumber
\begin{equation}
\label{eq:Xleft}
    (k_z a)_\mathrm{cr} = \frac{\pi \times \mbox{jet width}}{\mbox{jet height}},
\end{equation}
that implies one can speak for instability only if the normalized wavenumbers are larger than $(k_z a)_\mathrm{cr}$.  For $k_z a < (k_z a)_\mathrm{cr}$ one cannot talk for wave phenomena in general because the corresponding wavelength becomes larger than the jet's height.

Thus, KHI should occur in an instability window whose limits on the $k_z a$-axis are defined by aforementioned dimensionless $(k_z a)_\mathrm{cr}$ and $(k_z a)_\mathrm{rhs}$.  Those instability windows for the $m = 2$ and $m = 3$ MHD modes will be obtained in the next two subsections.  We note, that inequality (\ref{eq:instcond}), valid for untwisted rotating flux tube, can still be used in the case of a weakly twisted flux tube as such is modeled our rotating jet simply because the right-hand-side $k_z a$-value obtained from (\ref{eq:instcond}) is very close to the numerically found one for small magnitudes of the magnetic field twist parameter, say, $\varepsilon_1 = 0.001$, or even ${=}0.005$.  The usage of inequality (\ref{eq:instcond}) for weakly twisted flux tubes is not mandatory.  In principle, one can start searching the instability window from scratch---sooner or later one will find the right-hand-side limit of the instability window.  In any case (\ref{eq:instcond}) helps us to figure out approximately the range of dimensionless wavenumbers where the instability occurs, thus this formula has merely an ancillary character.

\subsection{Dispersion curves and growth rates of the unstable $m = 2$ MHD mode}
\label{subsec:m=2}
Let us first find what inequality, according to Eq.~(\ref{eq:instcond}), has to be satisfied in order to expect the possibility for instability occurrence.  The input parameters in that inequality for the mode number $m = 2$ are: $U_\phi = 136$~km\,s$^{-1}$, $v_\mathrm{Ai} \cong 150$~km\,s$^{-1}$, the density contrast $\eta = 0.474$, and the magnetic fields ratio $b = 1.27$ (bearing in mind that the environment is treated as cool medium).  Thus, one gets that the right-hand-side limit of the instability window is $(k_z a)_\mathrm{rhs} < 0.6456$.

The jet's height of $179$~Mm and width of $19.6$~Mm, according to expression (\ref{eq:Xleft}), define the left-hand-side limit of the instability window as $(k_z a)_\mathrm{cr} = 0.344$.  In other words, one can expect the occurrence of KHI within the instability window
\[
    0.344 \leqslant k_z a < 0.646.
\]

In solving the wave dispersion relation (\ref{eq:dispeq}) with aforementioned input values for $\eta$, $b$, and $\varepsilon_2$, we also assume that the magnetic flux tube is weakly twisted, choosing the magnetic field twist parameter $\varepsilon_1 = Aa/B_{\mathrm{i}z}$ to be equal to $0.005$.  Concerning the value of the Alfv\'en Mach number $M_\mathrm{A}$, we simply assume that the measured axial jet's speed $U_z = 114$~km\,s$^{-1}$ is the critical flow velocity at which the KHI starts.  This implies that (with $v_\mathrm{Ai} \cong 150$~km\,s$^{-1}$) the magnitude of $M_\mathrm{A}$ is equal to $0.76$.  It is worth pointing out,
however, that for a twisted magnetic field the magnitude of the axial magnetic field component $B_{\mathrm{i}z}$, which is used for computing both $\varepsilon_1$ and $v_\mathrm{Ai}$, is less than $B_{\mathrm{i}}$ itself.  This implies that the reference Alfv\'en speed $v_\mathrm{Ai}$ must be corrected along with the magnetic fields ratio $b$, notably $b$ should be replaced by $b_\mathrm{twist} = b\sqrt{1 + \varepsilon_1^2}$ and, accordingly, $v_\mathrm{Ai} \to v_\mathrm{Ai}/\sqrt{1 + \varepsilon_1^2}$ (see \citet{zhelyazkov2016}).  The diminished value of $v_\mathrm{Ai}$ requires a correction of the Alfv\'en Mach number, too,
\begin{figure}[!ht]
  \centering
\subfigure{\includegraphics[width = 2.6in]{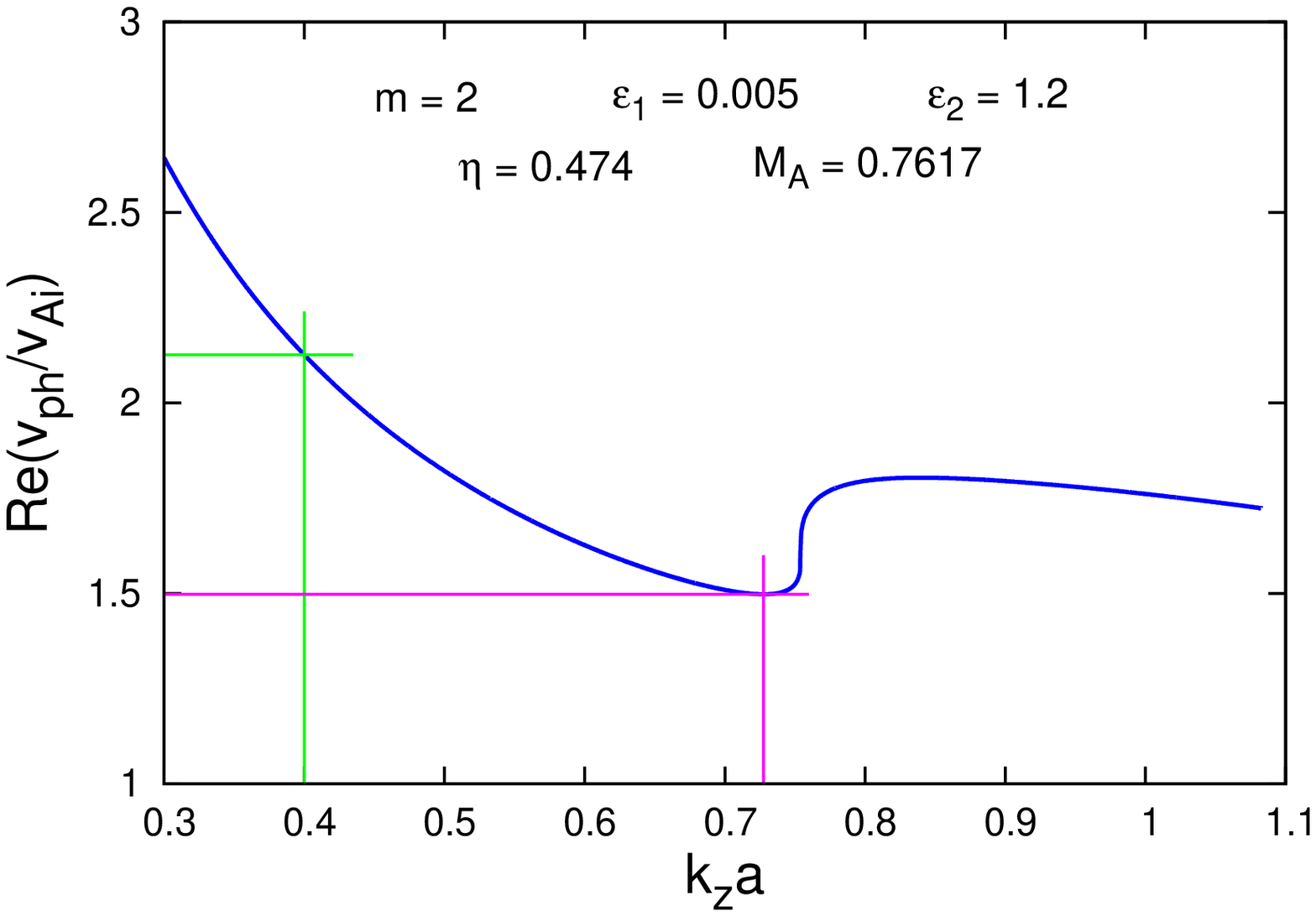}}
\subfigure{\includegraphics[width = 2.6in]{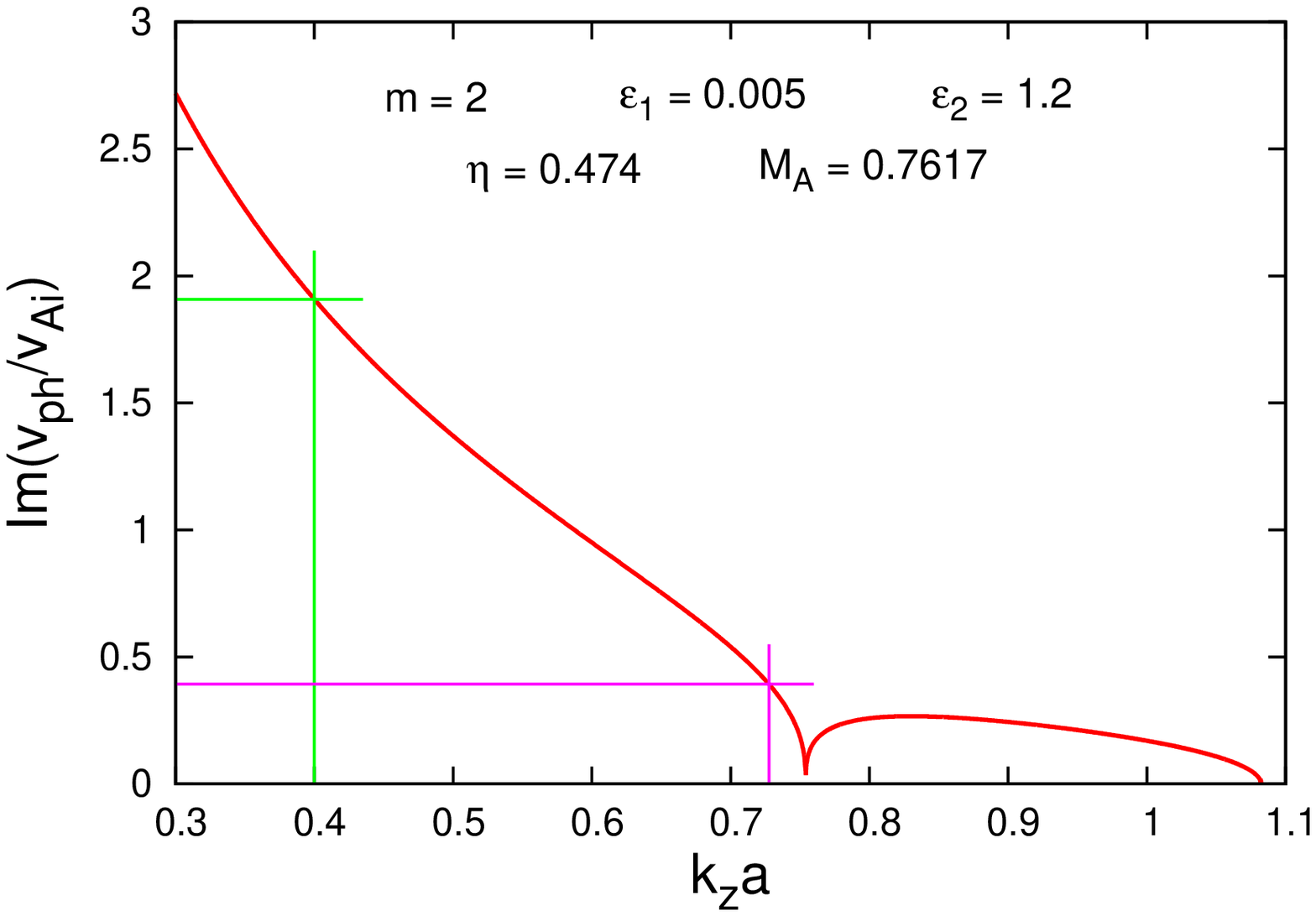}}
  \caption{(Left) Dispersion curve of unstable $m = 2$ mode in a twisted magnetic flux tube at $\eta = 0.474$, $b = 1.27$, $M_{\rm A} = 0.76$, $\varepsilon_1 = 0.005$, and $\varepsilon_2 = 1.2$.  (Right) Growth rates of the unstable $m = 3$ mode at the same input parameters.}
   \label{fig:fig3}
\end{figure}
namely $M_\mathrm{A}$ should be replaced by $M_\mathrm{A}\sqrt{1 + \varepsilon_1^2}$.  For $\varepsilon_1 = 0.005$ these corrections are negligibly small, thus we can take the aforementioned values of $b$ and $M_\mathrm{A}$ unchanged.   We note also that the $m = 2$ MHD mode in a rotating untwisted jet (that is, at $\varepsilon_1 = 0$) would become unstable within the same instability window---there should be no big difference between the dispersion and and growth rates curves calculated for these two values of $\varepsilon_1$.  Figure~\ref{fig:fig3} presents the results of the
numerical calculations for both the normalized wave phase velocity and growth rate as a function of the dimensionless wavenumber
$k_z a$. When observing the plots in Fig.~\ref{fig:fig3}, one immediately sees that both normalized
wave characteristics are of the same order in contrast, for instance, to the same plots for KHI in non-rotating jets, where the
normalized growth rates are,
in general, one order less than the normalized wave phase velocity.  The second distinctive issue is the complicated form of both curves, especially that for the growth rate.  The reason for such unexpected form of growth rate curve is the circumstance that at very small Alfv\'en Mach numbers one actually gets two different curves separated on the $k_z a$-axis.  One of them is gradually decreasing and has a cut-off at the approximately the predicted value for the right-hand-side limit $0.65$ of the instability window, while the other is a semi-closed ark-shaped curve with very small height being shifted to the right of the end of the first instability window.  With increasing $M_\mathrm{A}$, the small semi-closed curve moves to the left and at $M_\mathrm{A} = 0.76$ gently merges with the other curve; thus, one get the resulting growth rate curve which is situated in a wider instability window.

If we calculate the frequency growth rate and the instability wave phase velocity for two different wavelengths, determine by the corresponding choice of the dimensionless wave number [be it equal to $0.7276$ ($\lambda_\mathrm{KH} \cong 85$~Mm) or to $0.4$ ($\lambda_\mathrm{KH} = 154$~Mm)], one obtains the following values:
\[
    \gamma_\mathrm{KH} = 4.36 \times 10^{-3}\:\mathrm{s}^{-1}, \;\, v_\mathrm{ph} \cong 224\:\mathrm{km}\,\mathrm{s}^{-1} \;\,  \mbox{at}  \;\,  \lambda_\mathrm{KH} \cong 85~\mathrm{Mm}
\]
and
\[
    \gamma_\mathrm{KH} = 11.66 \times 10^{-3}\:\mathrm{s}^{-1}, \;\, v_\mathrm{ph} = 318.3\:\mathrm{km}\,\mathrm{s}^{-1}  \;\,  \mbox{at}  \;\,  \lambda_\mathrm{KH} = 154~\mathrm{Mm}.
\]

Having calculated frequency growth rate at fixed wavelengths, we can evaluate the time interval for the development of the KHI, $\tau_\mathrm{KH}$, by using the relation $\tau_\mathrm{KH} = 2\pi/\gamma_\mathrm{KH}$.  Thus, with $\varepsilon_1 = 0.005$, at the short wavelength of $85$~Mm, that time is $1440$~s or $24$~min, while at the longer wavelength of $154$~Mm this KHI developing time is $539$~s or ${\cong}9$~min.  Bearing in mind that the lifetime of the jet, as seen from Fig.~1 in \citet{chen2012}, is about $30$~min, then each KH unstable mode with growth time between $24$ and roughly $9$~min could in principle occur.

\citet{zaqarashvili2015} have derived the simple expression
\begin{equation}
\label{eq:tau}
    \frac{a}{U_\phi \sqrt{|m| - 1}}
\end{equation}
to estimate the growth time of KHI for different $m$ harmonics (supposing untwisted jets).  It is curious to compare the instability developing times calculated from this formula and obtained from a value of the frequency growth rate, $\gamma_\mathrm{KH}$, found for a small dimensionless wavenumber, $k_z a$.  The above expression, with $a = 9.8 \times 10^3$~km
and $U_\phi = 136$~km\,s$^{-1}$, yields an instability growing time of $72$~s or $1.2$~min.  It is interesting to note, that the frequency growth rate calculate for small normalized wavenumbers (in the range of $0.005$--$0.1$) and $b = 1$ has a very stable magnitude of $(13.24$--$13.39)\times 10^{-3}$~s$^{-1}$, which implies an instability developing time of (average) $472$~s or ${\cong}7.9$~min.  It seems that the above simple formula gives much shorter instability growing times, but recall that the same formula was derived from an expression for Im($\omega$) obtained in long-wavelength approximation, $k_z a \ll 1$, as well as in dense flux tube approximation, $\rho_\mathrm{e}/\rho_\mathrm{i} \equiv \eta \ll 1$.  Bearing in mind the fact that the jet's
radius and its height firmly fix [see Eq.~(\ref{eq:Xleft})] the left-hand-side limit of the instability window,
$(k_z a)_\mathrm{cr}$, a more realistic estimation of the KHI growth time can be obtained from the corresponding growth rate curve at a small $k_z a$, but lying within the instability window.

Instability criterion (\ref{eq:instcond}) does not depend on the magnetic twist parameter $\varepsilon_1 = Aa/B_{\mathrm{i}z}$, however, computations show that an increase in the value of $\varepsilon_1$ shifts to the left the right-hand-side limit of the
instability window.  This means that there should exist a critical $\varepsilon_1$ for which the right-hand-side limit of the
instability window coincides with $(k_z a)_\mathrm{cr}$.  In other words, a great enough magnetic field twist
\begin{figure}[!ht]
  \centering
\subfigure{\includegraphics[width = 2.6in]{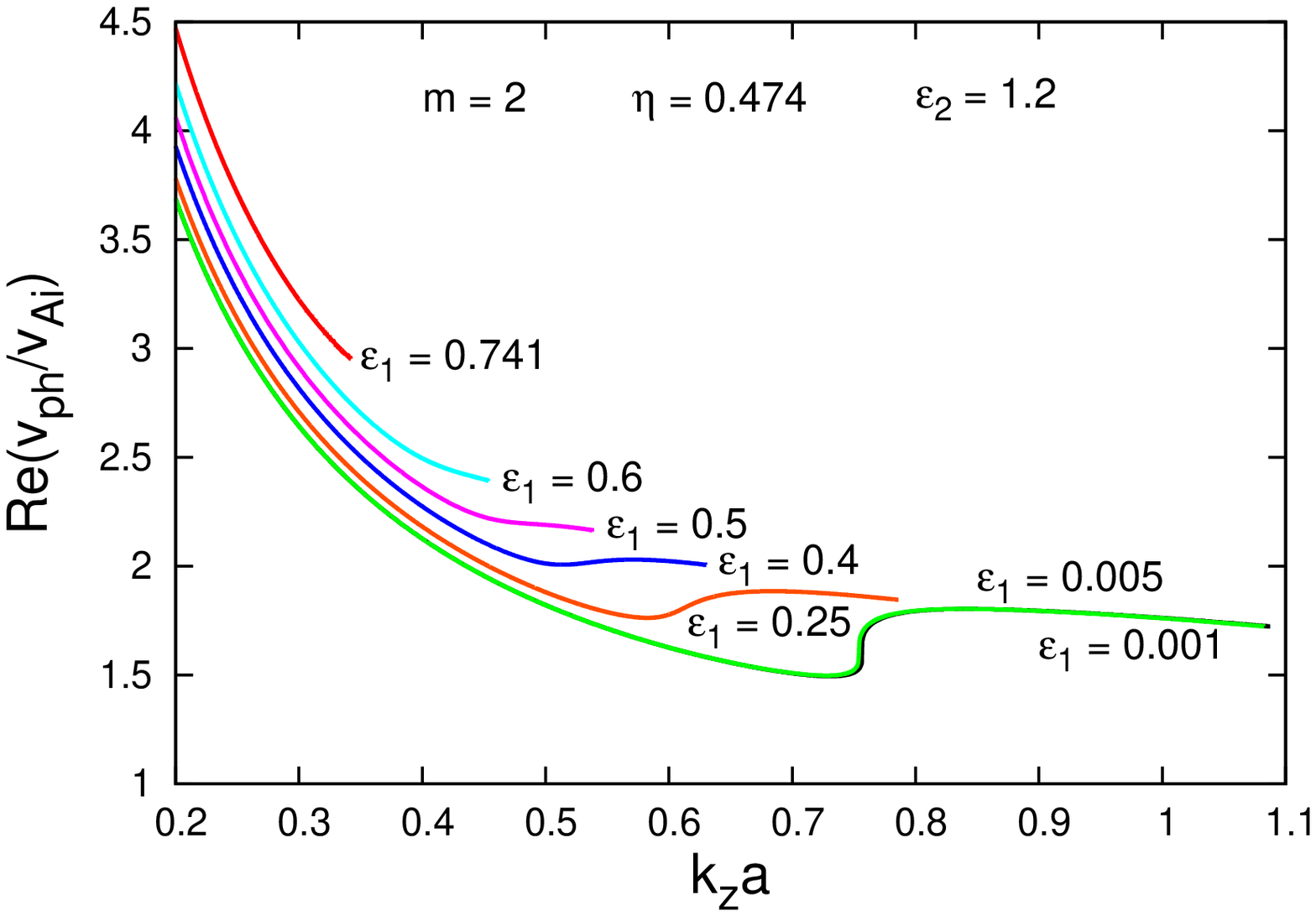}}
\subfigure{\includegraphics[width = 2.6in]{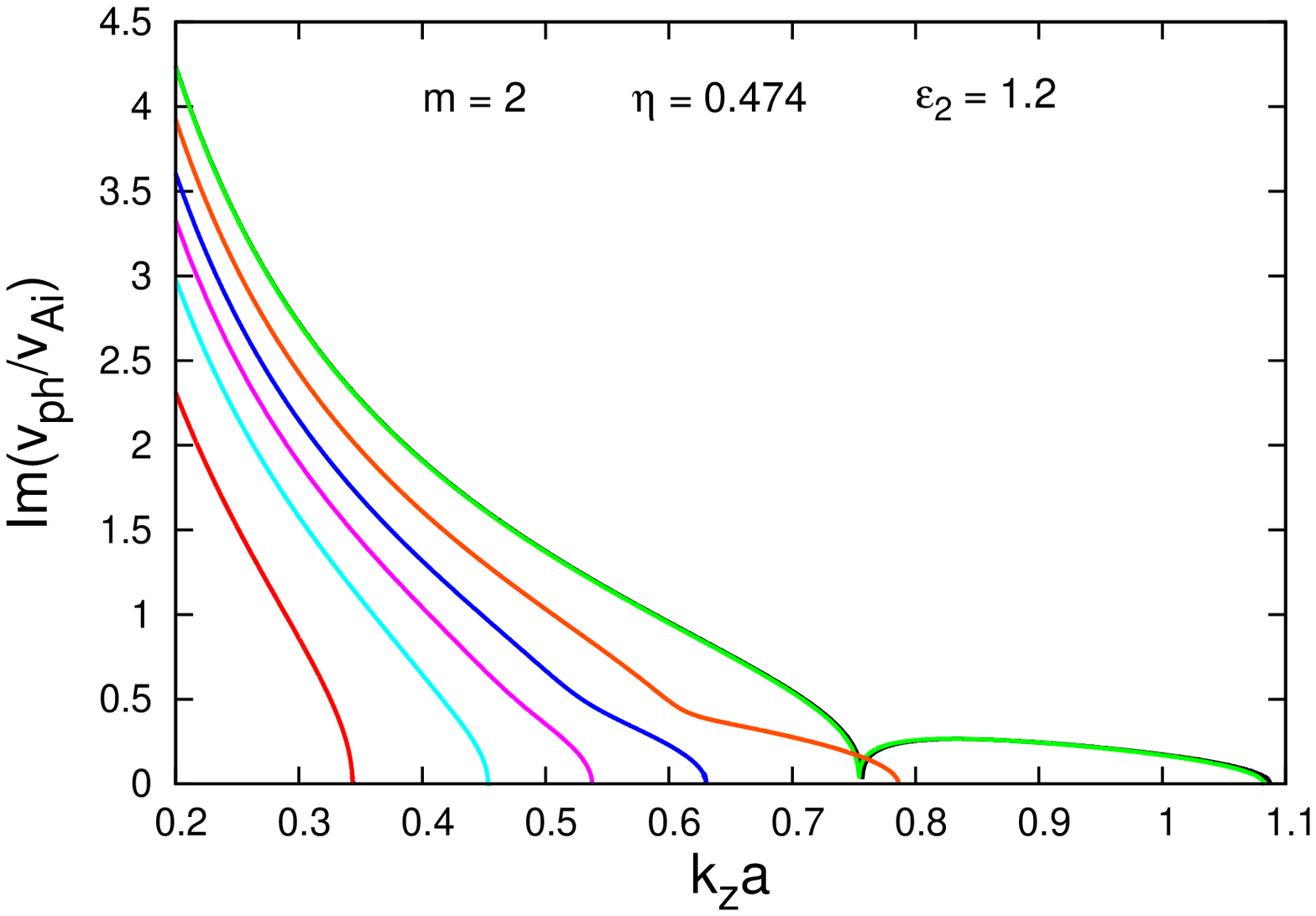}}
  \caption{(Left) Dispersion curves of unstable $m = 2$ MHD mode propagating along a twisted magnetic flux tube at $\eta = 0.474$, $\varepsilon_2 = 1.2$, and the following values of $\varepsilon_1$ (from right to left): $0.001$ (black curve), $0.005$, $0.25$, $0.4$, $0.5$, $0.6$, and $0.741$.  Alfv\'en Mach numbers for these curves are respectively $0.76$, $0.76$, $0.78$, $0.82$, $0.85$, $0.89$, and $0.9476$.  (Right) Growth rates of the unstable $m = 2$ mode for the same input parameters.  The azimuthal magnetic fields that corresponds to $\varepsilon_1 = 0.741$ (the instability window with zero width) is equal to $1.4$~G.  Real/observable $m = 2$ unstable MHD modes can be detected for $\varepsilon_1 < 0.741$, or, in other words, an azimuthal magnetic field of $1.4$~G would suppress the KHI onset.}
   \label{fig:fig4}
\end{figure}
can suppress the instability of the propagating MHD mode in the jet.  Figure~\ref{fig:fig4} shows the dependence of the instability window
on the value of magnetic field twist parameter $\varepsilon_1$, or equivalently on the magnetic field twist
$B_{\mathrm{i}\phi}(a) = \varepsilon_1  B_{\mathrm{i}z}$.  We note that for $\varepsilon_1 \geqslant 0.25$ one has to correct the values of $b$ and
$M_\mathrm{A}$ and as a result we have a series of progressively increasing values of these parameters starting from $\varepsilon_1 = 0.25$ onwards.  Moreover, it is clearly seen from the right panel of Fig.~\ref{fig:fig4} how the magnitude of the magnetic field twist parameter $\varepsilon_1$ controls the width of the instability window.  At large enough $\varepsilon_1 = 0.741$ (when $B_{\mathrm{i}z} \cong 1.9$~G) that width tends to zero.  In other words, an azimuthal magnetic field $B_{\mathrm{i}\phi}(a) = 1.4$~G is able to suppress KHI of the $m = 2$ MHD mode traveling on the jet.

Computations show that the width of the instability window and the typical frequency growth rates (or instability growing times), except of the magnetic field twist parameter $\varepsilon_1$, depend upon the azumuthal mode number $m$.  The same conclusion can be drawn from the instability criterion (\ref{eq:instcond}).  For example, at the same input values for $U_\phi$, $v_\mathrm{Ai}$, $\eta$, and $b$ as before, but with $m = 3$, the position of the right-hand-side limit of the instability window on the $k_z a$-axis is prescribed by $(k_z a)_\mathrm{rhs} = 1.018$.  Thus, now the width of the instability window is defined by the chain inequality $0.34 < k_z a < 1.02$.  As in the case of $m = 2$, we numerically obtained that: (i) the right-hand-side limit of the instability window lies far away from the predicted value of $1.0243$, notably around $1.6$, and (ii) both the dispersion and growth rate curves look smoother than the corresponding curves for $m = 2$.  Moreover, one obtains larger growth rates (and respectively shorter times for instability developing).  In the short-wavelength limit ($k_z a = 1.1905$, that is,
$\lambda_\mathrm{KH} \cong 52$~Mm) the instability growth rate is $\gamma_\mathrm{KH} = 7.73 \times 10^{-3}$~s$^{-1}$ which yields $\tau_\mathrm{KH} \equiv 2\pi/\gamma_\mathrm{KH} = 13.6$~min.  The wave phase velocity at this wavelength is equal to
$248.2$~km\,s$^{-1}$, i.e., it is slightly super-Alfv\'enic.  For the $m = 3$ MHD mode, the magnetic field twist parameter $\varepsilon_1$ at which the width of the instability window becomes equal to zero, has the magnitude of $0.765$ which implies that an azimuthal magnetic field (evaluated at the tube boundary with $B_{\mathrm{i}z} = 1.874$~G) of $1.43$~G would suppress the instability onset.

Although the $m = 3$ MHD mode can become unstable (at $\varepsilon_1 = 0.005$ and $k_z a = 1.1905$) with a $\lambda_\mathrm{KH} \cong 52$~Mm, this instability wavelength is still too long to fit the interspaces between growing blobs of the 2010 August 21
jet.  We do think that a wavelength in the range of $12$ to $15$~Mm will be suitable for studying the detected KHI in this case.
This implies, however, that we should jump to a higher $m$ MHD mode.  That said, dimensionless wave numbers in the range of
$4.1$--$5.13$ will fit the observations.  A rough evaluation of the $m$ that ensures aforementioned wavelengths at $U_\phi = 136$~km\,s$^{-1}$, $v_\mathrm{Ai} \cong 150$~km\,s$^{-1}$, a density contrast $\eta = 0.474$, magnetic fields ratio $b = 1.27$,  and $(k_z a)_\mathrm{rhs} = 5.5$ is obtained from inequality (\ref{eq:instcond}) and yields a value $m = 17$.  It turns out that this magnitude of the mode number $m$ is overestimated.  A smaller value, say, $m = 12$, is in general pretty good.  The fact that on using inequality (\ref{eq:instcond}) we obtained an overestimated value for $m$ is generally not surprising.  Recall that the
same inequality, in evaluating the expected instability windows' widths (more correctly their right-hand-side limits) for the $m = 2$ and $m = 3$ modes gave underestimated numbers---the computed right-hand-side limits turned out to be shifted to the right.  In the next subsection we present the dispersion and growth rate curves which we obtained for $m = 12$.

\subsection{Dispersion curves and growth rates of the unstable $m = 12$ MHD mode}
\label{subsec:m=12}
\begin{figure}[!ht]
  \centering
\subfigure{\includegraphics[width = 2.6in]{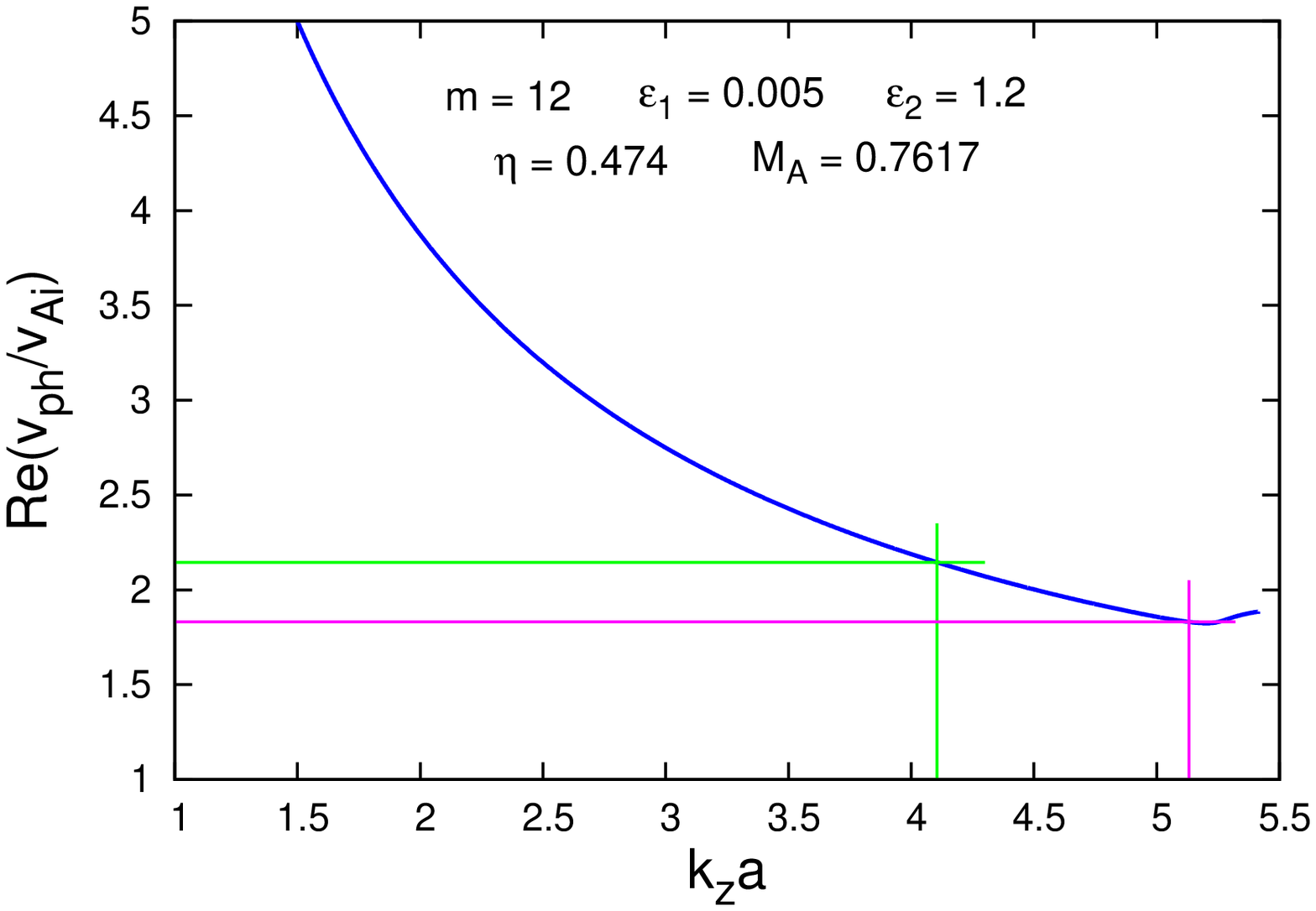}}
\subfigure{\includegraphics[width = 2.6in]{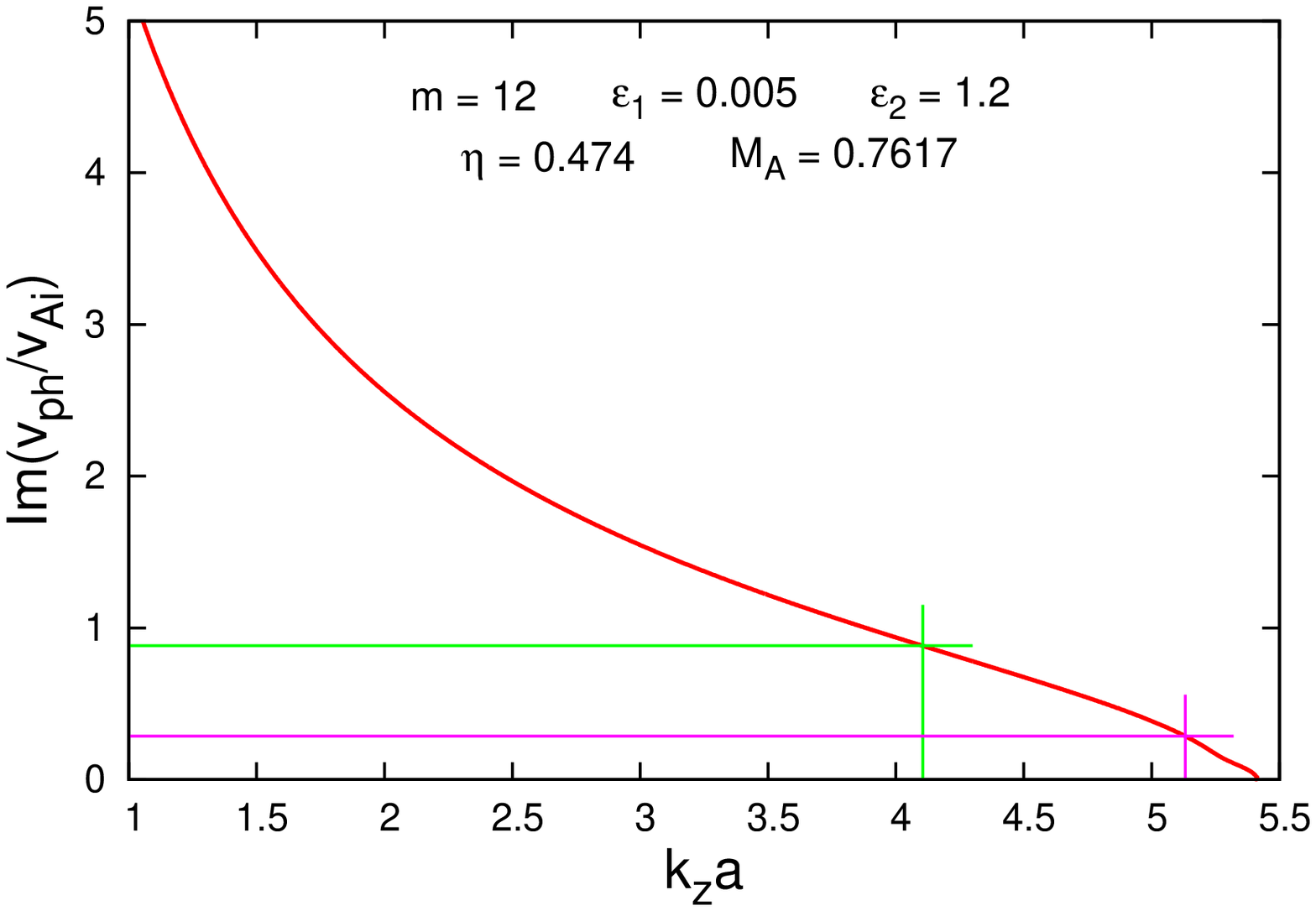}}
  \caption{(Left) Dispersion curve of unstable $m = 12$ mode in a twisted magnetic flux tube at $\eta = 0.474$, $b = 1.27$, $M_{\rm A} = 0.76$, $\varepsilon_1 = 0.005$, and $\varepsilon_2 = 1.2$.  (Right) Growth rates of the unstable $m = 12$ mode at the same input parameters.}
   \label{fig:fig5}
\end{figure}

The computations are straightforward and the results are presented in Fig.~\ref{fig:fig5} for instability wavelengths of
$\lambda_\mathrm{KH} = 15.0$ ($k_z a = 4.105$) and $12.0$~Mm ($k_z a = 5.131$), respectively.  The instability growth rate at $\lambda_\mathrm{KH} = 15.0$~Mm is equal to $\gamma_\mathrm{KH} \cong 55 \times 10^{-3}$~s, that yields an instability evolution time $\tau_\mathrm{KH} = 2\pi/\gamma_\mathrm{KH} = 114$~s $\cong 1.9$~min.  The mode phase velocity is equal to
${\cong}321$~km\,s$^{-1}$.  For the shorter instability wavelength of $12.0$~Mm the corresponding numbers are $\gamma_\mathrm{KH} \cong 22.44 \times 10^{-3}$~s, and $\tau_\mathrm{KH} = 280$~s $\cong 4.7$~min.  The wave phase velocity equals
\begin{figure}[!ht]
  \centering
\subfigure{\includegraphics[width = 2.6in]{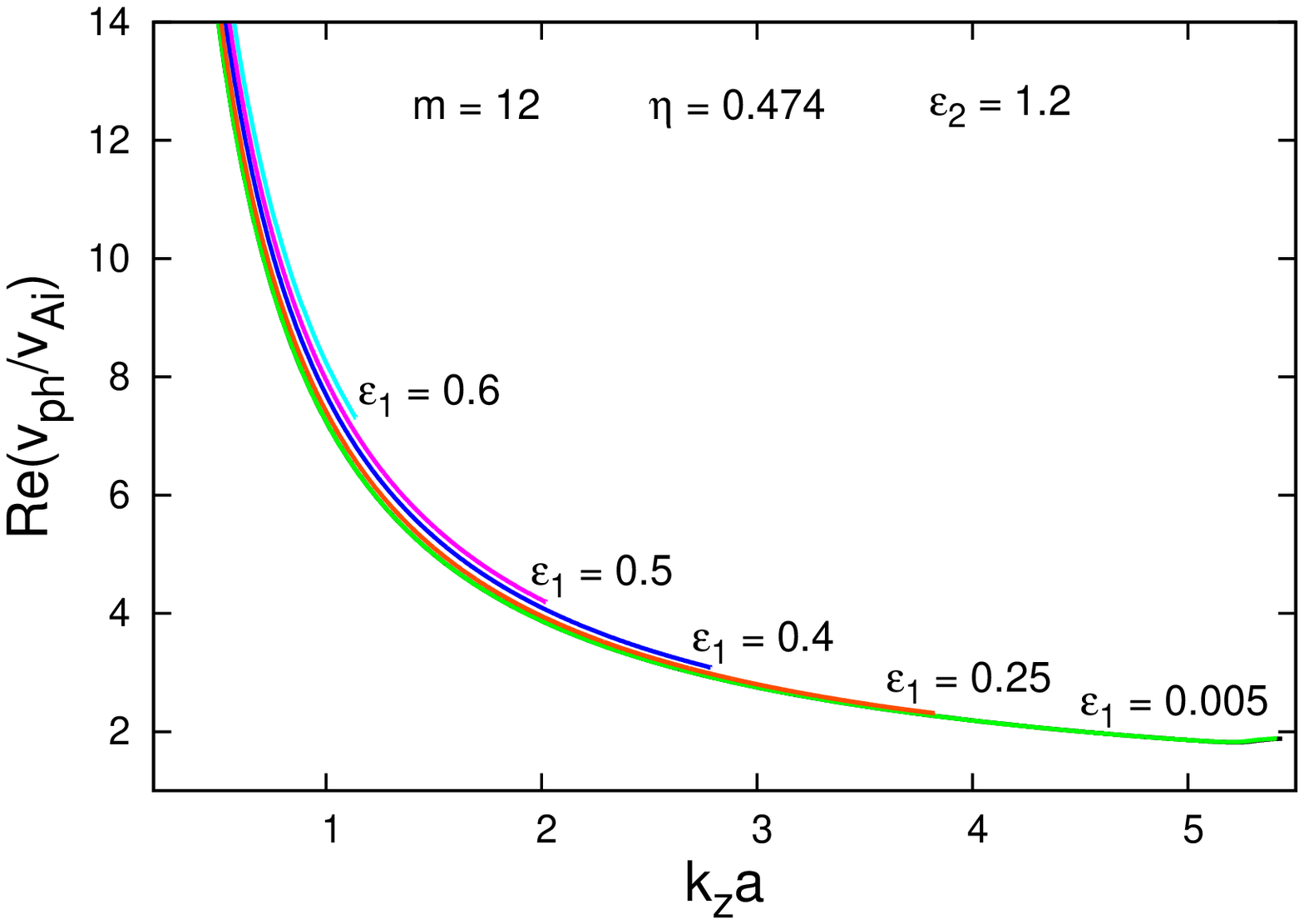}}
\subfigure{\includegraphics[width = 2.6in]{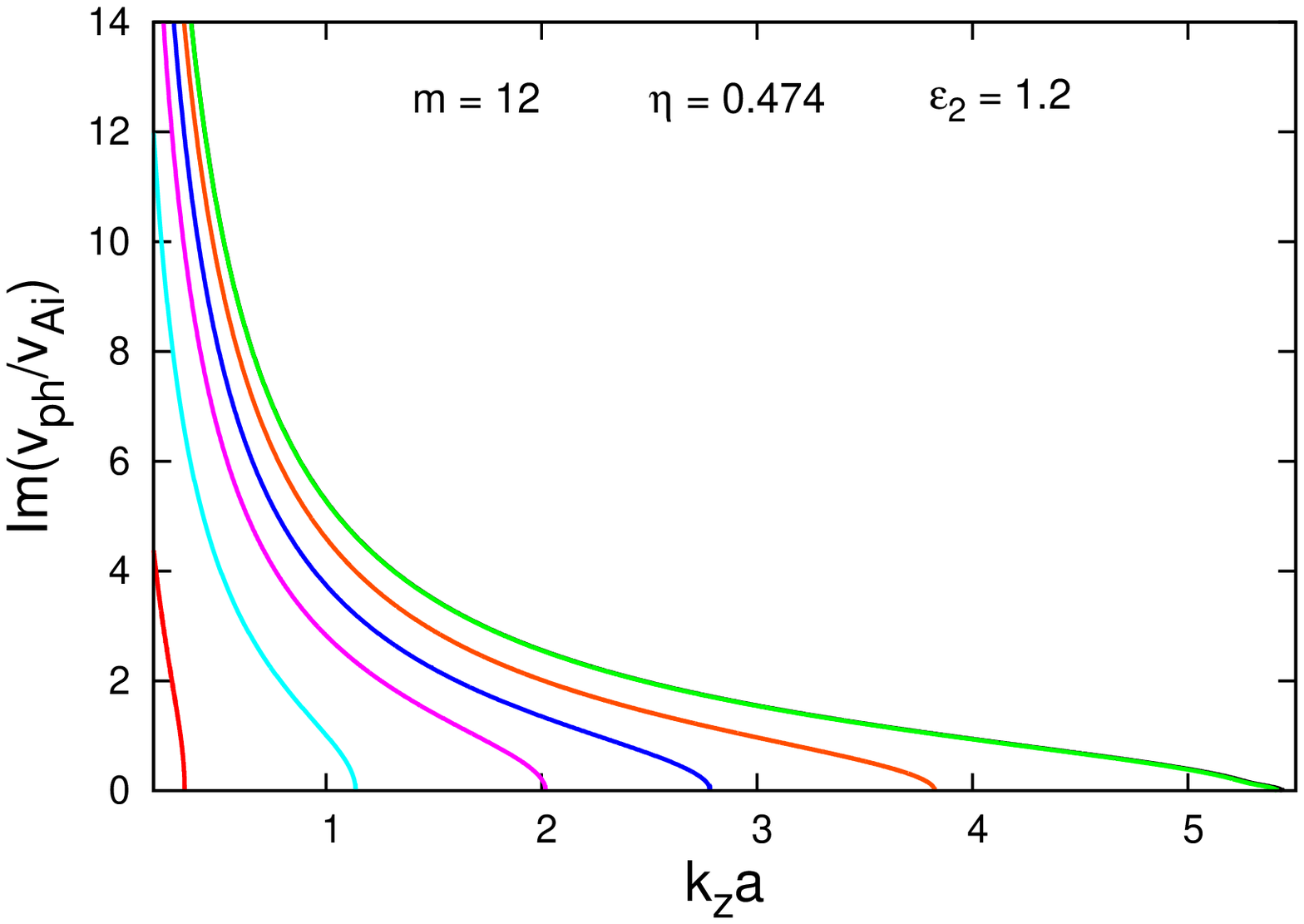}}
  \caption{(Left) Dispersion curves of unstable $m = 12$ MHD mode propagating along a twisted magnetic flux tube at $\eta = 0.474$, $\varepsilon_2 = 1.2$, and the following values of $\varepsilon_1$ (from right to left): $0.001$ (black curve, coinciding with the green one), $0.005$, $0.25$, $0.4$, $0.5$, $0.6$, and $0.67018$.  Alfv\'en Mach numbers for these curves are respectively $0.76$, $0.76$, $0.78$, $0.82$, $0.85$, $0.89$, and $0.9169$.  (Right) Growth rates of the unstable $m = 12$ mode for the same input parameters.  The azimuthal magnetic fields that corresponds to $\varepsilon_1 = 0.67018$ (the instability window with zero width) is equal to $1.3$~G.  Real/observable $m = 12$ unstable MHD modes can be detected for $\varepsilon_1 < 0.67018$, or, in other words, an azimuthal magnetic field of $1.3$~G would suppress the KHI onset.}
   \label{fig:fig6}
\end{figure}
${\cong}274$~km\,s$^{-1}$, that like the previous one is slightly super-Alfv\'enic.  We note that the KHI growth
rate could be estimated from the temporal evolution of the blobs in their initial stage and it is was found to be about $2$--$4$ minutes, so the numbers obtained from our plots are in qualitative agreement.

For the $m = 12$ MHD mode, the magnetic field twist parameter $\varepsilon_1$ at which the width of the instability window becomes equal to zero, has the magnitude of $0.67018$ (see red curve in the right panel of Fig.~\ref{fig:fig6}) which implies that an azimuthal magnetic field (evaluated at the tube boundary) of $1.3$~G would suppress the instability onset.  Recall that for the fluting ($m = 2$) and the $m = 3$  mode that magnetic field (calculated from correspondingly $\varepsilon_1 = 0.741$ and $0.765$) is slightly higher, namely equal to $1.4$ and $1.43$~G---the difference between the three magnetic fields is really small.

\section{Conclusion}
\label{sec:concl}
We have studied the development of KHI of the $m = 2,\;3$ and the $m = 12$ MHD modes traveling on rotating magnetically twisted moving flux tube of radius $a$ that models a tornado-like solar coronal hole jet near the northern pole of the Sun.  The spinning jet is treated as homogeneous magnetized incompressible plasma flux tube with a uniform density $\rho_\mathrm{i}$ surrounded by magnetized cool plasma with a constant density $\rho_\mathrm{e}$.  We consider a weakly twisted straight magnetic flux tube, therefore, it is stable against the kink \citep{zaqarashvili2014} and ballooning \citep{tsap2008} instabilities.  Twisted magnetic and velocity fields inside the jet are considered to be uniform characterized by the parameters $\varepsilon_1 = B_{\mathrm{i}\phi}(a)/B_{\mathrm{i}z}$ and $\varepsilon_2 = \Omega a/U_z$, where $\Omega$ is the angular velocity of the rotating jet and $U_z$ is its axial speed.  The magnetic field outside the jet is assumed to be homogeneous, that is, it is characterized by its magnitude $B_\mathrm{e}$ and is parallel to the jet's axial magnetic field.

The solution to the wave dispersion equation (in complex variables) for a given mode $|m| \geqslant 2$ requires $5$ input parameters, that can be obtained from the physical properties of the jet, namely the density contrast $\eta = \rho_\mathrm{e}/\rho_\mathrm{i}$, the magnetic fields' ratio $b = B_\mathrm{e}/B_{\mathrm{i}z}$, the two twist parameters $\varepsilon_1$ and $\varepsilon_2$, and Alfv\'en Mach number $M_\mathrm{A} = U_z/v_\mathrm{Ai}$, where $v_\mathrm{Ai} = B_{\mathrm{i}z}/\sqrt{\mu \rho_\mathrm{i}}$ is the Alfv\'en speed inside the jet.  In the present study we use the 2010 August 21 jet observed by \emph{SDO}/AIA \citep{chen2012} as a guide for the model parameters.

The numerical solution to dispersion equation shows that the propagating MHD mode can become unstable at Alfv\'en Mach numbers less than $1$ in an instability window on the $k_z a$-axis whose limits are specified by the mode number, $m$, jet's parameters, and by the ratio of the jet's width $\Delta \ell = 2a$ and jet's height.  As a rule, the higher mode number, the wider the instability window.  We note that KHI can arise even in rotating magnetically untwisted jets.  An increase in $\varepsilon_1$ leads to a narrowing of the instability parameter range and a big enough $\varepsilon_1$ can suppress KHI in the jet or, equivalently, there exists a critical azimuthal magnetic field $B_{\mathrm{i}\phi}(a)$ that can prevent the instability onset.

Wave growth rates and wave phase velocities of the unstable $m = 2,\;3$ and $m = 12$ modes depend on the wavelength: for instance, for the $m = 2$ MHD mode, at $\lambda_\mathrm{KH} \cong 85$~Mm the frequency growth rate is equal to $4.36 \times 10^{-3}$~s$^{-1}$, for the $m = 3$ mode at $\lambda_\mathrm{KH} \cong 52$~Mm is equal to $7.73 \times 10^{-3}$~s$^{-1}$, while for the $m = 12$ mode at $\lambda_\mathrm{KH} = 15.$~Mm it is equal to $55 \times 10^{-3}$~s$^{-1}$.  This difference in the frequency growth rates implies distinctive instability growth times whose values are $24$, $14$, and $1.9$ minutes of the three modes, respectively.  Guided by the observed distance between the blobs detected on the jet boundary we find that the $m = 12$ MHD mode can be associated with the observationally detected blobs instability.  Note that for the $m = 12$ MHD mode at wavelength of $12$~Mm the KHI developing time is $4.7$~min.  The phase velocities of unstable three modes are of the order of a few hundreds kilometers per second and the waves are super-Alfv\'enic.  It should be noted that we assume that observed blobs are results of KHI. However, other possibilities, such as non-uniformity in the background flow and magnetic field structures, as well as compressional wave modes, can not be ruled out.

KHI in the studied rotating and magnetically twisted jet can be suppressed by azimuthal magnetic fields of the order of $1.3$--$1.4$~G.  This circumstance might explain why the KHI is not detected in many observed events like solar coronal jets.  Another reason could be the insufficient resolution capabilities of observing instruments.

We should note that the modeling of KHI in the unwinding polar jet, observed by \citet{shen2011}, would give similar results for the instability growth rates (or instability developing times) and the phase velocities of the unstable MHD modes as these obtained in our study, because jet's speeds and other data in \citet{shen2011} are similar to those of \citet{chen2012}.  A fine tuning might require a change of the high mode number $m$ from $12$ to $13$, or to $11$, but this is not a problem at all.

In any case, our approach of exploring KHI in rotating jets is applicable for any observationally detected instability provided we know the above-mentioned physical parameters of the jet, and in particular, the critical axial flow velocity $U_z^{\mathrm{cr}}$ for instability onset along with the rotational speed $U_\phi$ and the observationally deduced wavelength $\lambda_\mathrm{KH}$.  Estimated from observations instability growth rate and wave phase velocity, compared with their values obtained from numerically derived growth rate and dispersion curves, can serve as a benchmark for the applicability of our approach in studying the KHI in tornado-like solar jets.  The example with the $m = 12$ MHD mode shows that our model is rather flexible and for given observational data of detected/visualized KHI one can find out the appropriate mode number that fits well the observations.  The main limitations of our model are the linear analysis and the compressibility.  The nonlinearity leads to the saturation of the growth of KHI, and formation of nonlinear waves \citep{miura1982,miura1984}.  Compressibility as well may affect the KHI, and may lead to somewhat reduced growth rate when the density contrast is high \citep{ofman2011}.

\vspace{4.4mm}

\textbf{Acknowledgments} \hspace{0.5mm}
The work of I.Zh.\ and R.C.\ was supported by the Bulgarian Science Fund and the Department of Science \& Technology, Government of India Fund under Indo-Bulgarian bilateral project DNTS/INDIA 01/7, /Int/Bulga\-ria/P-2/12.  The work of T.V.Z.\ was supported by the Austrian Fonds zur F\"{o}rderung der Wissenschaftlichen Forschung under projects P26181-N27 and P25640-N27, and by
the Georgian Shota Rustaveli National Science Foundation project DI-2016-17.  L.O.\ would like to acknowledge support by NASA grant NNX16AF78G.  The authors are indebted to the two anonymous reviewers for their helpful and constructive comments that greatly contributed to improving the final version of the manuscript.

\end{document}